\documentclass[useAMS,usenatbib]{mn2e}
\usepackage{lscape,graphicx}
\usepackage{rotating}
\usepackage{color}

\usepackage{color}

\newcommand{\ewr}{\mbox{$W_r(2796)$}}

\newcommand{\lya}{\mbox{${\rm Ly}\alpha$}}

\newcommand{\apg}{\:^{>}_{\sim}\:}
\newcommand{\apl}{\:^{<}_{\sim}\:}
\newcommand{\etal}{\ensuremath{\mbox{et~al.}}}

\providecommand{\kms}{\,\ensuremath{\rm{km\,s}^{-1}}}

\newcommand\mnras{{MNRAS}}%
\newcommand\araa{{ARA\&A}}%
\newcommand\apj{{ApJ}}%
\newcommand\apjl{{ApJ}}%
\newcommand\apjs{{ApJS}}%
\newcommand\apss{{Ap\&SS}}%
\newcommand\aap{{A\&A}}%
\title[Empirical Constraints of Super-Galactic Winds at $z\apg 0.5$]{Empirical Constraints of Super-Galactic Winds at $z\apg 0.5$}

\author[Gauthier \& Chen]{Jean-Ren\'e Gauthier$^1$\thanks{E-mail:jrg@astro.caltech.edu} and  Hsiao-Wen Chen$^2$\thanks{E-mail:hchen@oddjob.uchicago.edu} \\
\\
$^{1}${Cahill Center for Astronomy and Astrophysics, California Institute of Technology, Pasadena CA 91125 USA }\\
$^{2}${Department of Astronomy \& Astrophysics and Kavli Institute for Cosmological Physics, University of Chicago, Chicago IL 60637 USA} }

\pagerange{\pageref{firstpage}--\pageref{lastpage}} \pubyear{2012}

\begin{document}

\label{firstpage}

\maketitle

\begin{abstract}

  Under the hypothesis that Mg\,II absorbers found near the minor axis
  of a disk galaxy originate in the cool phase of super-galactic
  winds, we carry out a study to constrain the properties of
  large-scale galactic outflows at redshift $z_{\rm gal}\apg 0.5$
  based on the observed relative motions of individual absorbing
  clouds with respect to the positions and orientations of the
  absorbing galaxies.  We identify in the literature four highly
  inclined disk galaxies located within 50 kpc and with the minor axis
  oriented within 45 degrees of a background QSO sightline. Deep HST
  images of the galaxies are available for accurate characterizations
  of the optical morphologies of the galaxies.  High-quality echelle
  spectra of the QSO members are also available in public archives for
  resolving the velocity field of individual absorption clumps.  Three
  galaxies in our sample are located at $\rho=8-34$ kpc and exhibit
  strong associated Mg\,II absorption feature with $\ewr \apg 0.8$
  \AA.  One galaxy, located at an impact parameters $\rho=48$ kpc, dose
  not show an associated Mg\,II absorber to a 3-$\sigma$ limit of
  $\ewr=0.01$ \AA.  Combining known morphological parameters of the
  galaxies such as the inclination and orientation angles of the
  star-forming disks, and resolved absorption profiles of the
  associated absorbers at $\rho < 35$ kpc away, we explore the allowed
  parameter space for the opening angle $\theta_0$ and the velocity
  field of large-scale galactic outflows as a function of $z$-height,
  $v(z)$.  We find that the observed absorption profiles of the Mg\,II
  doublets and their associated Fe\,II series are compatible with the
  absorbing gas being either accelerated or decelerated, depending on
  $\theta_0$, though accelerated outflows are a valid characterization
  only for a narrow range of $\theta_0$.  Under an acceleration
  scenario, we compare the derived $v(z)$ with predictions from Murray
  \etal\ (2011) and find that if the gas is being accelerated by the
  radiation and ram pressure forces from super star clusters, then the
  efficiency of thermal energy input from a supernova explosion is
  $\epsilon \apl 0.01$.  In addition, we adopt a power-law function
  from Steidel \etal\ (2010) for characterizing the accelerated
  outflows as a function of $z$-height, $a(z)\propto z^{-\alpha}$.  We
  find a steep slope of $\alpha\approx 3$ for a launch radius of
  $z_{\rm min}=1$ kpc.  A shallower slope of $\alpha\approx 1.5$ would
  increase $z_{\rm min}$ to beyond 4 kpc.  We discuss the implications
  of these parameter constraints.

\end{abstract}



\section{Introduction}

Super-galactic winds are considered a promising mechanism to eject
metals into the circumgalactic and intergalactic media (e.g.,
\citealt{aguirre2001a,cen2005a,oppenheimer2006a,pieri2007a,dave2010a}).
Galactic-scale outflows are observed through a variety of techniques
and diagnostics, including detections of blueshifted self-absorption
in Na\,I $\lambda\lambda$ 5890,5896 (e.g.,
\citealt{heckman2000a,rupke2005a, rupke2005b}) or in Mg\,II
$\lambda\lambda 2796,2803$ doublet transitions (e.g.,
\citealt{tremonti2007a,weiner2009a}; Martin \& Bouch\'e 2009; Rubin et
al.\ 2010), and resolved morphologies of hot plasma associated with
supernova driven winds (e.g., \citealt{veilleux2005a}).  While the
observed outflow speed extends up to $600-1000$ \kms\ (e.g.,
\citealt{rupke2005a, rupke2005b, weiner2009a}), starburst driven winds
have not been directly detected beyond $\sim 15$ kpc from nearby
star-forming regions (e.g., Heckman et al.\ 2002; Rubin et al.\ 2011).

At the same time, observations of close QSO--galaxy and galaxy--galaxy
pairs have revealed chemically enriched circumgalactic medium out to
100 kpc projected distances through absorption features imprinted in
the spectra of background objects (e.g., Chen et al.\ 2001a;
Adelberger et al.\ 2005; Chen et al.\ 2010a; Steidel et al.\ 2010)
Although there is a lack of observational evidence to physically
connect outflows that produce blueshifted self-absorption in
star-forming galaxies with metal enriched absorbers found at $10-100$
kpc along transverse directions, starburst driven outflows remain as
the leading scenario for explaining the presence of metal enriched
absorbing clouds at large projected distances (e.g., Oppenheimer et
al.\ 2009).  In particular, many groups have attempted to connect
strong metal-line absorbers such as Mg\,II of rest-frame absorption
equivalent width $\ewr\apg 1$ \AA\ with star-forming galaxies (e.g.,
Bond \etal\ 2001; Zibetti et al.\ 2007; Chelouche \& Bowen 2010;
Kacprzak \& Churchill 2011; Nestor \etal\ 2011; Matejek \& Simcoe
2012;
but see Gauthier \& Chen 2011;  
Chen et al.\ 2010b for different empirical findings).  
A key result of these various studies is that star-forming galaxies at
redshift $z_{\rm gal}\sim 1$ appear to show enhanced Mg\,II
absorption\footnote{The use of Mg\,II absorbers is mostly practical,
  because at $z\apg 0.4$ the doublet transitions are redshifted into
  the optical window where QSO and galaxy spectra are recorded and
  because these transitions are strong.} at projected distances $\apl
50$ kpc along the minor axis (e.g., Bordoloi et al.\ 2011), suggesting
that outflows may be a dominant contributor to the observed Mg\,II
absorbers at least within 50 kpc of star-forming regions.

Recent analytic works have examined different physical mechanisms that
could drive super-galactic winds (e.g.\,
\citealt{murray2005a,murray2011a}).  In \citet{murray2011a}, the
authors propose that radiation pressure from massive star clusters can
drive outflows at velocities greater than the galaxy circular
velocity.  Once the outflowing material reaches above the star-forming
disk, it is exposed to both radiation from other star-forming regions
and hot gas from supernova bubbles originated in the parent galaxy.
According to this model, the combination of radiation pressure and ram
pressure force can accelerate and drive outflows to $\sim 50-100$ kpc
away.  Such mechanisms can be particularly efficient in low-mass
galaxies undergoing strong episodes of star formation activities like
M82.

Under the hypothesis that Mg\,II absorbers originate in super-galactic
winds, empirical constraints for the geometry and dynamics of outflows
can be obtained based on comparisons of line-of-sight gas kinematics
and spatial orientation of the star-forming disk.  Specifically, in
the local universe large-scale galactic outflows are commonly found to
follow a bi-conical pattern along the rotation axis of the
star-forming disk (e.g., \citealt{heckman2000a}) with a varying degree
of collimation, typically $10-45^{\circ}$ \citep{bland-hawthorn2007a}.
In this paper, we test the superwind hypothesis of the origin of
Mg\,II absorbers and derive constraints for starburst driven outflows
using pairs of QSOs and highly inclined disk galaxies found at $z_{\rm
  gal}=0.2-0.9$.  We have identified in the literature four highly
inclined disk galaxies located within 50 kpc and with the minor axis
oriented within 45 degrees of a background QSO sightline.  Deep {\it
  Hubble Space Telescope} (HST) images of the galaxies are available
in the HST data archive for accurate characterizations of the optical
morphologies of the galaxies.  High-quality echelle spectra of the QSO
members are available in public archives for resolving the velocity
field of individual absorption clumps.  We construct a conical outflow
model and apply the observed gas kinematics to constrain the velocity
gradient along the polar axis of each galaxy.  We compare the results
of our analysis with different model predictions (e.g., Steidel \etal\
2010; Murray \etal\ 2011) and identify a plausible range for various
model parameters.

This paper is organized as follows. In section 2, we describe the data
and observational techniques.  A description of the conic outflow
model for describing the wind geometry is presented in section 3.  In
section 4, we present the derived constraints of the outflow model for
individual galaxies.  We compare the results of our study with
predictions of analytical models in section 5.  We adopt a $\Lambda$
cosmology with $\Omega_M = 0.3$ and $\Omega_{\Lambda} = 0.7$, and a
Hubble constant $H_0 =70$ \kms Mpc$^{-1}$ throughout the paper. All
distances are in proper rather than comoving units unless otherwise
stated.

\begin{footnotesize}
\begin{table*}
  \centering
  \begin{minipage}{160mm}
    \caption{Properties of Sample Galaxies and the Associated Mg\,II Absorbers }
    \begin{tabular}{@{}lcrrcrrrccrc@{}}
      \hline 
      &  &  & \multicolumn{1}{c}{$\rho$}  &  & \multicolumn{1}{c}{$i_0$}    &   \multicolumn{1}{c}{$\alpha$}  & $W_r([{\rm O\,II}])$ &  &  & & $W_r(2796)$$^c$ \\ 
      \multicolumn{1}{c}{QSO}  & Galaxy  &  \multicolumn{1}{c}{$z_{\rm gal}$}  & (kpc) & $M_B$ & \multicolumn{1}{c}{($^{\circ}$)} & \multicolumn{1}{c}{($^{\circ}$)} &     \multicolumn{1}{c}{(\AA)}   &  Ref$^{\,a}$    &  &  $z_{\rm Mg\,II}$$^b$ &    (\AA)  \\ 
      \hline
      \hline
      3C336         & A & 0.4721 & 33.6 & -19.0 & 74 & 96.8 &  & 1,3 & & 0.4715 & 0.82$\pm$0.07  \\ 
      & B & 0.8909 & 23.3 & -20.3 & 81 & 124.1 & $\approx -13.4$ & 1,3 & & 0.8907 & 1.53$\pm$0.05 \\ 
      LBQS0058$+$0155 & C & 0.6120 &  7.9 & -18.4 & 65  & 113 & $-2.6\pm1.1$ & 3,5 & & 0.6126 & 1.67$\pm$0.01  \\ 
      Q0827$+$243     & D &  0.199 & 48.4 & -19.9 & 85 & 86.3 &  & 4 & &  0.199     & $< 0.01$ \\ 
      \hline
      \multicolumn{12}{l}{$^a$[1] Steidel et al.\ (1997); [2] Chen et al. (2001a); [3] Chen et al.\ (2005); [4] Steidel et al.\ (2002); [5] Pettini et al.\ (2000).} \\
      \multicolumn{12}{l}{$^b$We adopt a mean absorption redshift as the redshift of the absorber.} \\
      \multicolumn{12}{l}{$^c$Total rest-frame absorption equivalent width summed over all components.} 
    \label{pair_table}
  \end{tabular}
\end{minipage}
\end{table*}
\end{footnotesize}

\section{Observational Data and Analysis}

To obtain empirical constraints for starburst driven outflows, we
searched the literature and identified a small sample of close
(projected distances $\rho<50$ kpc) QSO and galaxy pairs for which
the disks have an inclination angle
$i_0>60$ degrees and a position angle of the major axis $\alpha>
45$ degrees (or minor axis $<45$ degrees) from the QSO sightline.
These selection criteria ensure that the QSO sightline probes the 
regions near the polar axis. 


Our search yielded four QSO and galaxy pairs. 
A summary of the physical properties of the galaxies are
presented in Table \ref{pair_table}, including galaxy redshift
($z_{\rm gal}$), projected distance ($\rho$) to the QSO sightline,
rest-frame absolute $B$-band magnitude ($M_B$), inclination ($i_0$)
and position angles ($\alpha$) of the star-forming disk and, when
available, the rest-frame equivalent width of [O\,II] emission line
($W_r({\rm [O\,II]})$).  Unfortunately, no measurement of on-going
star formation rate is available for any of the galaxies in the
sample.  We also present in Table 1 the properties of the
corresponding Mg\,II absorber, including the mean absorption redshift
($z_{\rm Mg\,II}$) and total rest-frame absorption equivalent width
($\ewr$).  In the absence of a Mg\,II absorption feature, we place a
3-$\sigma$ upper limit to $\ewr$.

Deep optical HST images obtained using the Wide Field and Planetary
Camera 2 (WFPC2) and the F702W filter are available for all four
galaxies in the HST data archive.  Morphological parameters of
galaxies $A$ and $B$ in the field around 3C336 were adopted from Chen
et al.\ (1998, 2001b).  Galaxy $C$ in the field around LBQS0058$+$0155
is heavily blended with the background QSO light.  We adopted the
inclination angle estimated by Pettini et al.\ (2000) and determine
the position angle of the inclined disk ourselves.  Galaxy $D$ in the
field around Q0827$+$243 has not been studied before.  We retrieved
the imaging data and determined the morphological parameters by
performing a two-dimensional surface brightness profile analysis
ourselves.

Echelle spectra of 3C336 and Q0827$+$243 were obtained using the
Ultraviolet and Visual Echelle Spectrograph (UVES;
\citet{dodorico2000a}) on VLT-UT2 and was retrieved from the ESO data
archive. The observations of 3C336 (program ID 069.A-0371, PI:
Savaglio) and Q0827$+$243 (program ID 068.A-0170) were carried out
using a $1''$ slit, dichroic \#1, and cross-disperser CD\#1, yielding
a spectral resolution of ${\rm FWHM}\,\approx 7.5$ \kms.  A sequence
of three exposures of 4900 s each were obtained for 3C336 and a
sequence of four exposures of 3600 s duration were obtained for
Q0827$+$243. The data were binned 2$\times$2 during readout.  The
spectra were processed and calibrated using the standard UVES
reduction pipeline.

Echelle spectra of LBQS$0058+0155$ were retrieved from the Keck HIRES
KOA archive.  The observations of LBQS$0058+0155$ (program ID C99H,
PI: Steidel) were obtained using a $0.9''$ slit (Deck C1) and the UV
cross-disperser that offer a spectral resolution of ${\rm
  FWHM}\,\approx 6$ \kms.  A sequence of two exposures of 3600 s each
were taken.  The data were binned 2$\times$2 during readout.  The
spectra were processed through the MAKEE reduction pipeline, a
reduction code tailored to handle HIRES data
\footnote{http://spider.ipac.caltech.edu/staff/tab/makee/}.  The
product of MAKEE consists of individual echelle orders, corrected for
the blaze function, vacuum wavelengths, and heliocentric velocities.

Individual spectra were combined to form a final stacked spectrum per
QSO, excluding deviant pixels, which was then continuum normalized
based on a low-order polynomial fit to regions in the QSO spectrum
that are free of narrow absorption lines.  
These final combined spectra were used to examine the kinematic
signatures of Mg\,II absorbers and those of associated metal-line
transitions such as Fe\,II, and Mn\,II absorption at the redshift of
each galaxy listed in Table 1.  The high spectral resolution and high
$S/N$ QSO spectra allow us to resolve individual absorption components
and determine accurate velocity centroids of these individual clumps.

\section{A Conical Outflow Model}

Under the hypothesis that strong Mg\,II absorbers uncovered near the
polar axis of a star-forming galaxy are produced in super-galactic
winds, here we develop a simple conical outflow model to characterize
the geometry and kinematics of the outflowing gas.  Specifically, we
assume that the observed Mg\,II absorbers originate in the
photo-ionized, cool phase (with temperature $T\sim 10^4$ K) of conical
outflows which follow the path of least resistance along the polar
axis.

\begin{figure*}
\centerline{
\includegraphics[angle=0,scale=0.5]{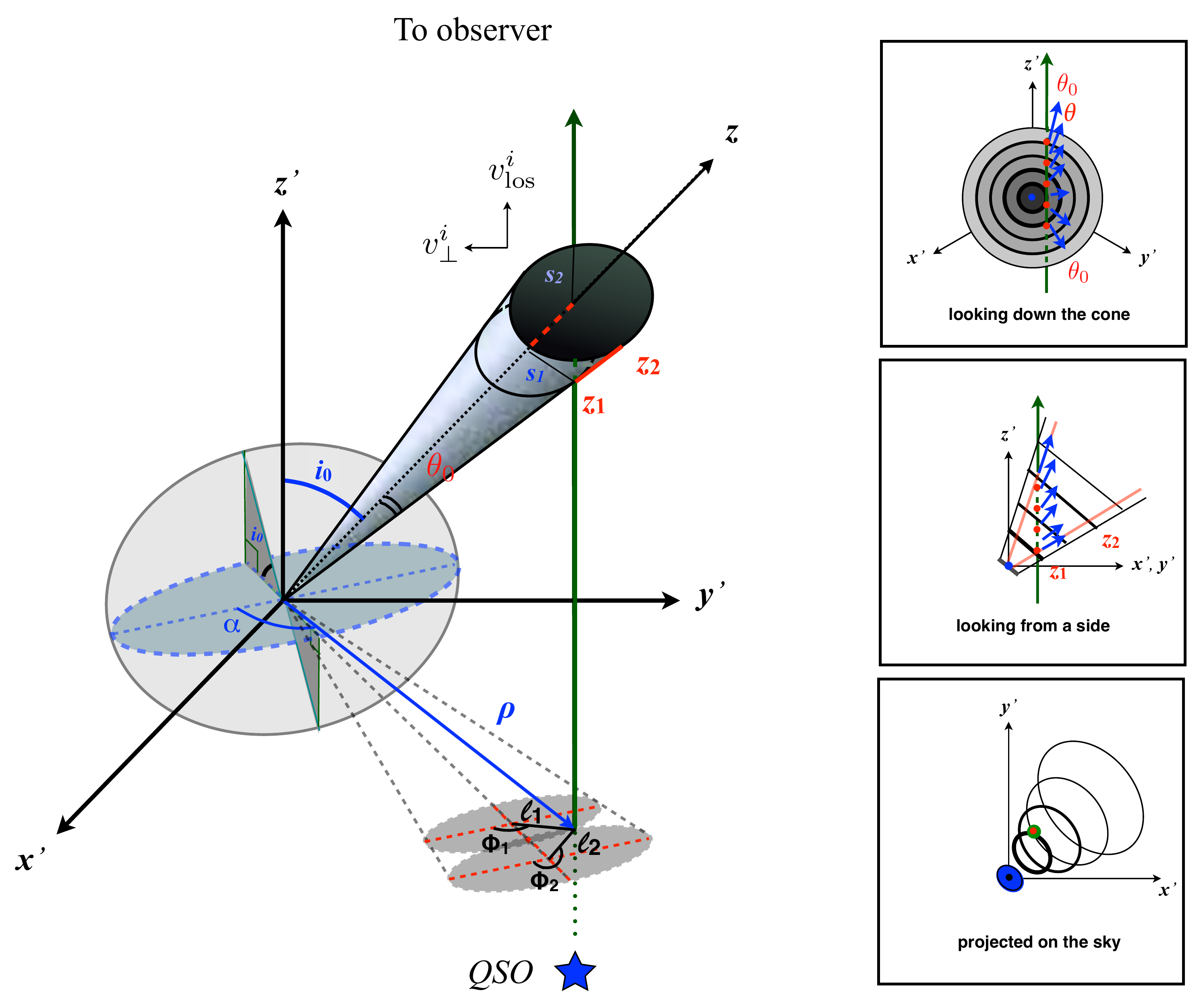}
}
\caption{{\it Left}: Cartoon illustrating the conical outflow model in
  our study.  In this viewing angle, the QSO line of sight runs in
  parallel to the $\vec{\bf z'}$ axis, while the $x'$-$y'$ plane
  represents the plane of the sky.  The star-forming disk is indicated
  by the grey circle and is inclined from the $\vec{\bf z'}$ axis by
  $i_0$ degrees.  The projected disk on the plane of the sky is shown
  by the light blue ellipse oriented at a position angle of $\alpha$
  from the QSO sightline, which is at projected distance $\rho$.  The
  conical outflows follow the path of least resistance along the polar
  axis ($\vec{\bf z}$) with a total angular span of $2\,\theta_0$.
  The solid ellipses along the conical outflows mark the disks at
  constant $z$-height in the outflows, with their projection on the
  plane of the sky indicated by the corresponding light grey ellipses
  in the $x'$-$y'$ plane.  The QSO line of sight enters the outflows
  at $z_1$ and exits at $z_2$ ($> z_1$).  The size of the outflowing
  disk at $z_{[1,2]}$ is $s_{[1,2]}=z_{[1,2]}\,\tan\theta_0$.  The
  projected distances between the outflow disks at $z_{[1,2]}$ and the
  QSO sightline are marked by $\ell_{[1,2]}$, which are related to
  $z_{[1,2]}$ according to
  $s_{[1,2]}=\ell_{[1,2]}\,\sqrt{1+\sin^2\phi_{[1,2]}\,\tan^2
    i_0}=z_{[1,2]}\,\tan\theta_0$.  {\it Right}: Viewing the conical
  outflows in different projections.  The panels show the impact
  geometry of the QSO sightline when looking straight down the
  collimated outflows (top), when looking from a side (middle), and
  when projected on the sky (bottom).  The QSO line of sight is marked
  in green with the path inside the outflows highlighted in red dots.
  The blue arrows indicate the moving direction of the outflowing gas
  at an angle $\theta$ from $\vec{\bf z}$.}
\label{cartoon}
\end{figure*}

The conical outflow model is illustrated in Figure \ref{cartoon},
which shows the relative orientations of the inclined disk and the
outflows with respect to the QSO line of sight.  The collimated
outflow, originated at the center of the star-forming disk, is
characterized by an outward expanding cone along the polar axis
$\vec{\bf z}$ with a total angular span of $2\,\theta_0$, and the size
of the outflowing disk $s$ at $z$ is $s=z\,\tan\theta_0$.  The
star-forming disk (at $z=0$) marked by the grey circle is inclined
from the line of sight $\vec{\bf z'}$ by $i_0$ degrees.  The
projection of the star-forming disk on the plane of the sky (the
$x'$-$y'$ plane) is shown in light blue with the major axis oriented
at a position angle $\alpha$ from the QSO line of sight, which is at
projected distance $\rho$.  The QSO sightline intercepts the conical
outflows at a vertical disk height $z$ from $z_1$ to $z_2$, which is
determined by the opening angle $\theta_0$.  The projected distances
$\ell_{[1,2]}$ between the outflow disks at $z_{[1,2]}$ and the QSO
sightline are related to $z_{[1,2]}$ according to
$s_{[1,2]}=\ell_{[1,2]}\,\sqrt{1+\sin^2\phi_{[1,2]}\,\tan^2
  i_0}=z_{[1,2]}\,\tan\theta_0$, where $\phi_{[1,2]}$ are the position
angles of the projected outflow disks at $z_{[1,2]}$ (see Figure
\ref{cartoon}) and are constrained by
\begin{equation}
\tan\phi_{[1,2]}=\frac{\rho\,\sin\alpha-z_{[1,2]}\,\sin i_0}{\rho\,\cos\alpha}.
\end{equation}
Applying the law of sines, which relates $\rho$ and $\ell$ following
\begin{equation}
\frac{\rho}{\sin(3\,\pi/2-\phi)} = \frac{\ell}{\sin(\alpha-\pi/2)}
\end{equation} 
the relation between ($z_1$,$z_2$) and the opening angle $\theta_0$ is
then formulated according to the following,
\begin{equation}
z_{[1,2]}\,\tan\theta_0=\rho\sqrt{1+\sin^2\phi_{[1,2]}\tan^2i_0}\,\left(\frac{\cos\alpha}{\cos\phi_{[1,2]}}\right),
\end{equation}
where $i_0$ and $\alpha$ are the inclination and orientation angles of
the star-forming disk.

Equations (1) and (3) can be generalized to calculate the appropriate
$\theta$ for any given point along the QSO line of sight at $z$-height
$z_1\le z\le z_2$ (the top-right panel of Figure \ref{cartoon}).  For
an absorbing clump moving outward at $z$ height, the outflow speed $v$
is then related to the observed line-of-sight velocity $v_{\rm los}$
according to
\begin{equation}
v = v_{\,\rm los}\,/\cos i
\end{equation}
where 
\begin{equation}
i = \sin^{-1}\left(\frac{\rho}{z}\cos\theta\right),
\end{equation}
and $\theta \le \theta_0$.

For our study, $i_0$, $\alpha$, and $\rho$ are known from HST imaging
data of the galaxies, and $\theta_0$, which is unknown, uniquely
determines the minimum and maximum $z$-height, $z_1$ and $z_2$ in the
outflows as probed by the QSO sightline.  Here we assume that the cool
gas producing the observed absorption features is distributed
symmetrically around the polar axis and the absorbing clumps at $z_1$
and $z_2$ probe regions close to the front and back side of the
conical outflows.  This is consistent with the outflow morphologies
seen in local starburst galaxies (see Cooper et al. 2008 for a more
detailed discussion).  The goal of our study is to explore a plausible
range of $\theta_0$ and examine how the velocity of outflowing gas
varies with $z$-height.  We note that if asymmetry arises due to
inhomogeneities in the outflows, then the inferred velocity gradient
(see \S\ 5.1) represents a lower limit to the intrinsic outflow
velocity field.

\section{Empirical Constraints of Individual Galaxies}

Using the geometric model developed in \S\ 3 for conical outflows, we
proceed with kinematic studies of individual galaxies.  For each
galaxy, we take into account its known optical morphology from HST
images, which determines the orientation of the conical outflows.  We
measure the line-of-sight velocity field of the outflows based on the
resolved absorption components at the redshift of the galaxy observed
in the echelle spectra of the background QSO.

To measure the velocity field, we first search in the QSO echelle
spectra for all absorption transitions that are associated with the
galaxy.  These include, in addition to the Mg\,II absorption doublets,
the Fe\,II absorption series, Mn\,II, and Mg\,I absorption
transitions.  We then perform a Voigt profile analysis that considers
all the observed absorption features at once, using the
VPFIT\footnote{http://www.ast.cam.ac.uk/~rfc/vpfit.html} software
package (Carswell \etal\ 1991).  We consider the minimum number of
components required to deliver the best $\chi^2$ in the Voigt profile
analysis.  Finally, we establish the observed line-of-sight velocity
in the outflows by comparing the relative velocities of individual
absorption component with the systemic redshift of the galaxy.

\begin{figure}
\centerline{
\includegraphics[angle=0,scale=0.40]{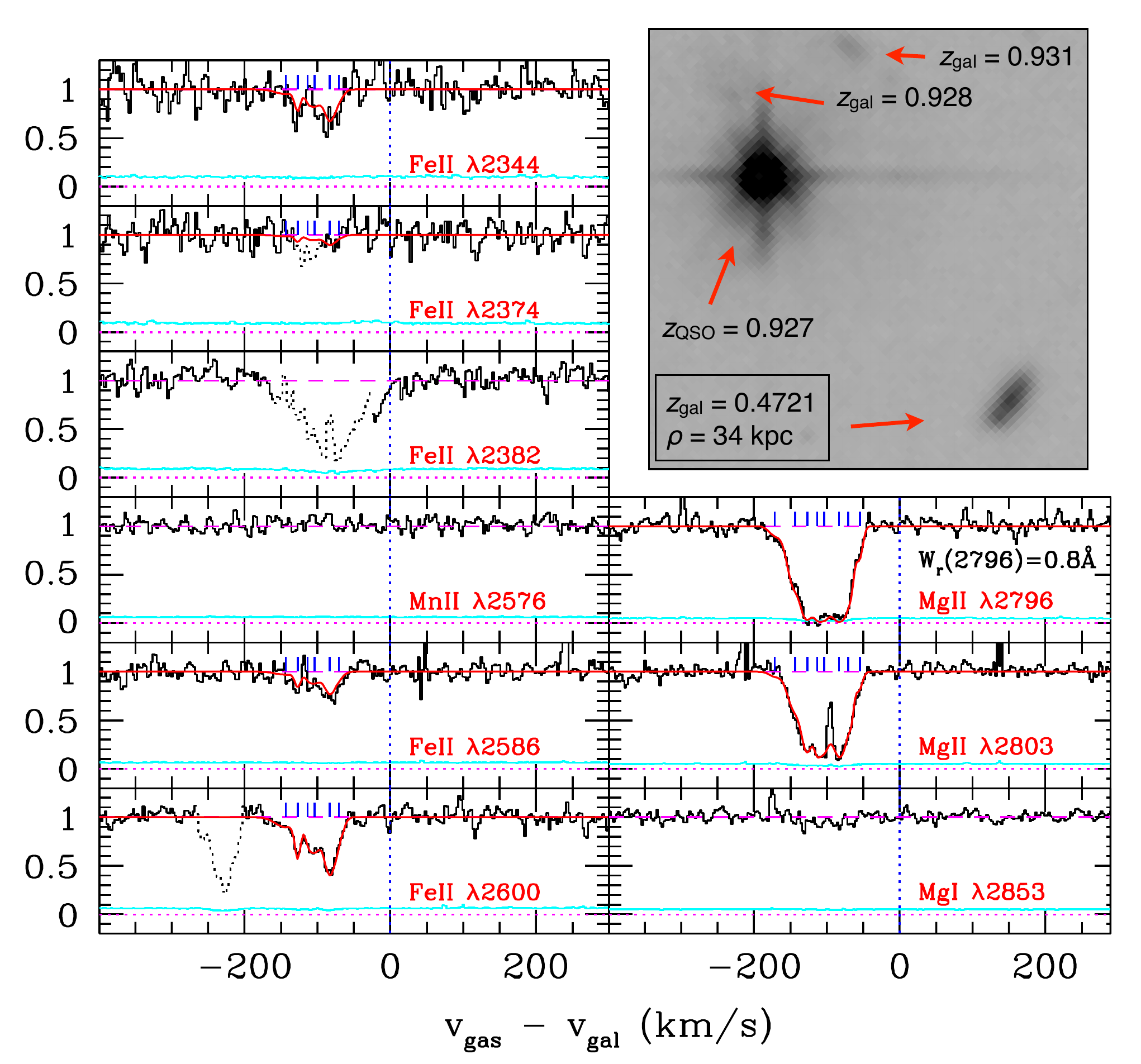}
}
\caption{Line-of-sight velocity distribution of absorbing clouds at
  projected distance $\rho=34$ kpc of Galaxy A at $z_{\rm gal}=0.472$
  with $\alpha=96.8$ degrees and $i_0=74$ degrees.  The QSO echelle
  spectra cover absorption transitions due to Fe\,II, Mn\,II, Mg\,II,
  and Mg\,I at the redshift of the galaxy.  The absorption profiles
  are shown in individual spectral panels.  We observe strong
  absorption in Fe\,II and Mg\,II, but not in Mn\,II or Mg\,I
  transitions.  In each spectral panel, the absorption spectrum is
  shown in solid histogram with the 1-$\sigma$ error spectrum shown in
  thin cyan histograms.  Zero velocity in each spectral panel
  corresponds to the systemic redshift of the galaxy at $z_{\rm
    gal}=0.4721$.  Contaminating features are dotted out.  A Voigt
  profile analysis of the observed Fe\,II and Mg\,II absorption
  profiles yields a minimum of eight individual absorption components
  and a reduced $\chi^2$ of $\chi^2_r=1.1$.  The best-fit model
  absorption profiles are shown in red curves and the positions of
  individual components are also marked by tickmarks at the top of
  individual panels.  The total rest-frame Mg\,II absorption
  equivalent width over all observed components is $\ewr=0.8$ \AA.
  The absorbing clumps display relative line-of-sight motions ranging
  from $\Delta\,v_{\rm los}=-54.8$ \kms\ to $\Delta\,v_{\rm
    los}=-143.7$ \kms\ with respect to the systemic redshift of the
  galaxy.  The absorbing galaxy is located in the lower-right corner
  of the image panel.  The background QSO and additional
  spectroscopically identified galaxies are also marked.}
\label{3C336_z04715}
\end{figure}

\begin{figure}
\centerline{
\includegraphics[angle=0,scale=0.50]{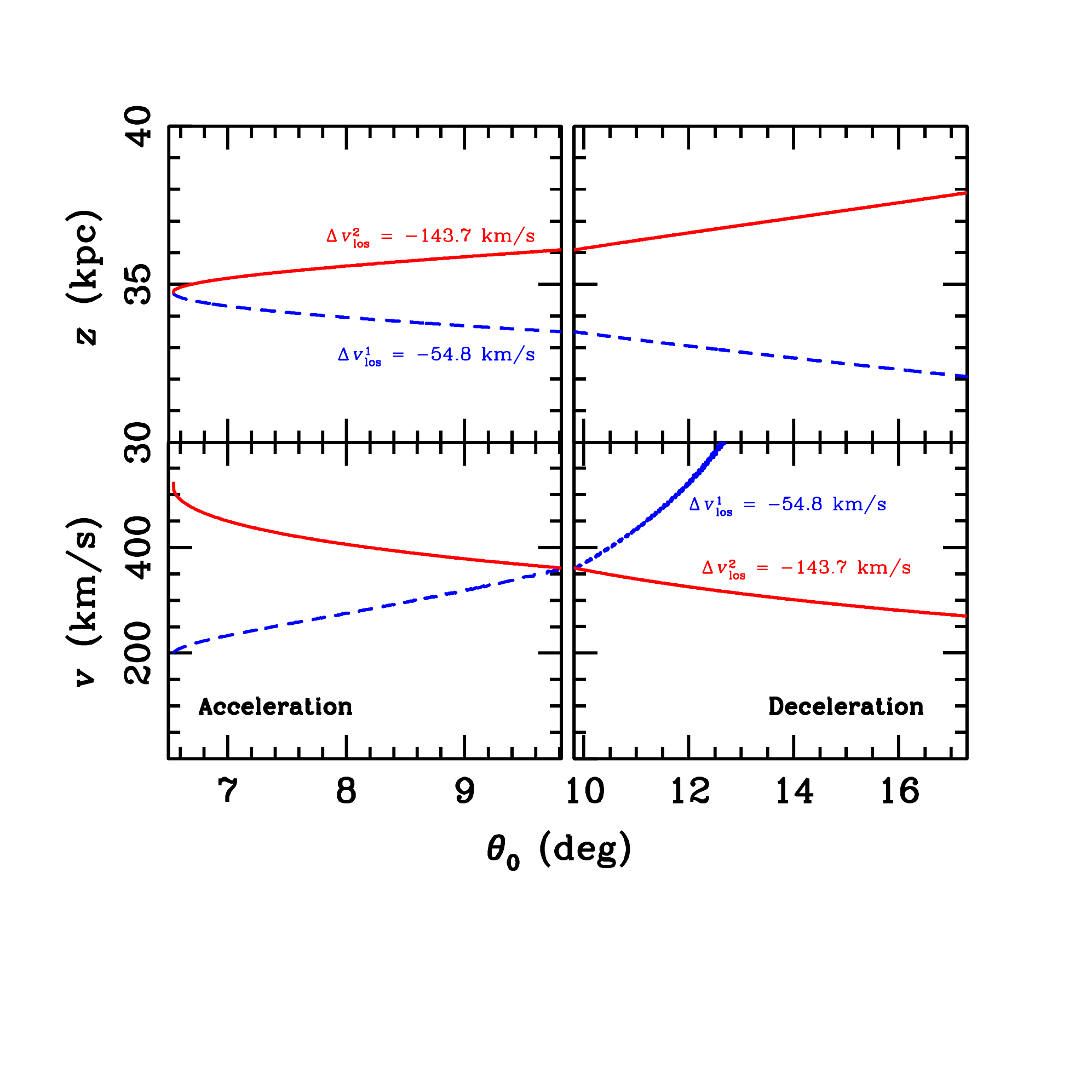}
}
\caption{Allowed parameter space for the $z$-heights (top panels) and
  de-projected velocities (bottom) of individual absorbing components
  observed in Figure \ref{3C336_z04715} versus allowed opening angle
  $\theta_0$.  The minimum and maximum allowed $\theta_0$ are
  constrained by the relative orientation and alignment of the
  star-forming disk with respect to the QSO sightline.  As shown in
  Figure \ref{3C336_z04715}, the galaxy is oriented at a position
  angle of $\alpha=96.8$ degrees from the QSO line of sight.  In order
  for outflows to be responsible for the observed absorption features
  in the QSO spectrum, the minimum allowed opening angle is $\theta_0
  \apg 6.5$ degrees.  In addition, all absorbing clumps are found
  blueshifted from the systemic redshift of the galaxy.  Given that
  the galaxy has a inclination angle of $i_0=74$ degrees, the lack of
  redshifted absorbing components constrains the opening angle at
  $\theta_0\apl 16$ degrees.  If we further assume that the
  line-of-sight velocity increases smoothly from $z_1$ (where the QSO
  sightline enters the conical outflows) to $z_2$ (where the QSO
  sightline exits the conical outflows), we find that the outflows can
  only be accelerating if $\theta_0\apl 10$ degrees.  Beyond
  $\theta_0\approx 10$ degrees, the absorbing clump at $z_1$ would
  have to move faster that the absorbing clump at $z_2$ in order to
  produce the observed line-of-sight velocity of $\Delta\,v_{\rm
    los}^1=-54.8$ \kms, in which case the outflows would be
  decelerating as the gas moves further away from the star-forming
  disk. }
\label{thvza_1_z04715}
\end{figure}

Assuming that the outflow velocity field is characterized by a smooth
velocity gradient with the distance from the star-forming disk, we can
then obtain a unique mapping between the observed velocity components
and the corresponding $z$-heights in the conical outflow model.  The
observed relative motions of individual absorbing clumps along the QSO
sightline, when deprojected along the polar axis according to
Equations (4) and (5) for a plausible range of $\theta_0$, constrain
the outflow velocity field as a function of $z$-height, $v(z)$, that
can be compared directly with model predictions (e.g., Martin \&
Bouch\'e 2009; Steidel \etal\ 2010; Murray \etal\ 2011).  We describe
the results of our analysis of individual galaxies in the following
sections.

\subsection{Galaxy A at $z=0.472$ in the field around 3C~336}

Galaxy A in the field aroud 3C~336 ($z_{\rm QSO} =0.927$) was
spectroscopically identifed at $z_{\rm gal}=0.4721\pm 0.0002$ by
\citet{steidel1997a}.  The galaxy is at projected distance $\rho=33.6$
kpc from the QSO line of sight.  Chen \etal\ (1998, 2001b) analyzed
available HST WFPC2 images of the field (top-right panel of Figure 2)
and measured $\alpha=96.8$ degrees and $i_0=74$ degrees for the disk
(Table 1).  The echelle spectra of the QSO cover a wavelength range
that allows observations of Fe\,II, Mn\,II, Mg\,II, and Mg\,I
absorption at the redshift of the galaxy.  The absorption profiles are
shown in individual spectral panels of Figure 2.  We detect strong
absorption complex in Fe\,II and Mg\,II, but not in Mn\,II or Mg\,I
transitions.  A Voigt profile analysis that simultaneously takes into
account the observed Fe\,II and Mg\,II absorption profiles yields a
minimum of eight individual absorption components and $\chi^2_r=1.1$.
The total rest-frame Mg\,II absorption equivalent width over all
observed components is $\ewr=0.8\pm 0.1$ \AA.  The absorbing clumps
display relative line-of-sight motions ranging from $\Delta\,v_{\rm
  los}=-54.8$ \kms\ to $\Delta\,v_{\rm los}=-143.7$ \kms\ with respect
to the systemic redshift of the galaxy.

Figure 2 shows that the cool gas probed by the Mg\,II absorption
transitions is entirely blueshifted with respect to the star-forming
disk with a total line-of-sight velocity spread of $\approx
90$\kms. Given that the galaxy is oriented at a position angle of
$\alpha=96.8$ degrees from the QSO line of sight, we note that in
order for outflows to be responsible for the observed absorption
features in the QSO spectrum the opening angle must exceed $\theta_0
\approx 6.5$ degrees.  Furthermore, the galaxy has a inclination angle
of $i_0=74$ degrees, and therefore the lack of redshifted absorbing
components constrains the opening angle at $\theta_0\apl 16$ degrees.
The minimum and maximum allowed $\theta_0$ are completely constrained
by the relative orientation and alignment of the star-forming disk
with respect to the QSO sightline.

\begin{figure}
\centerline{
\includegraphics[angle=0,scale=0.40]{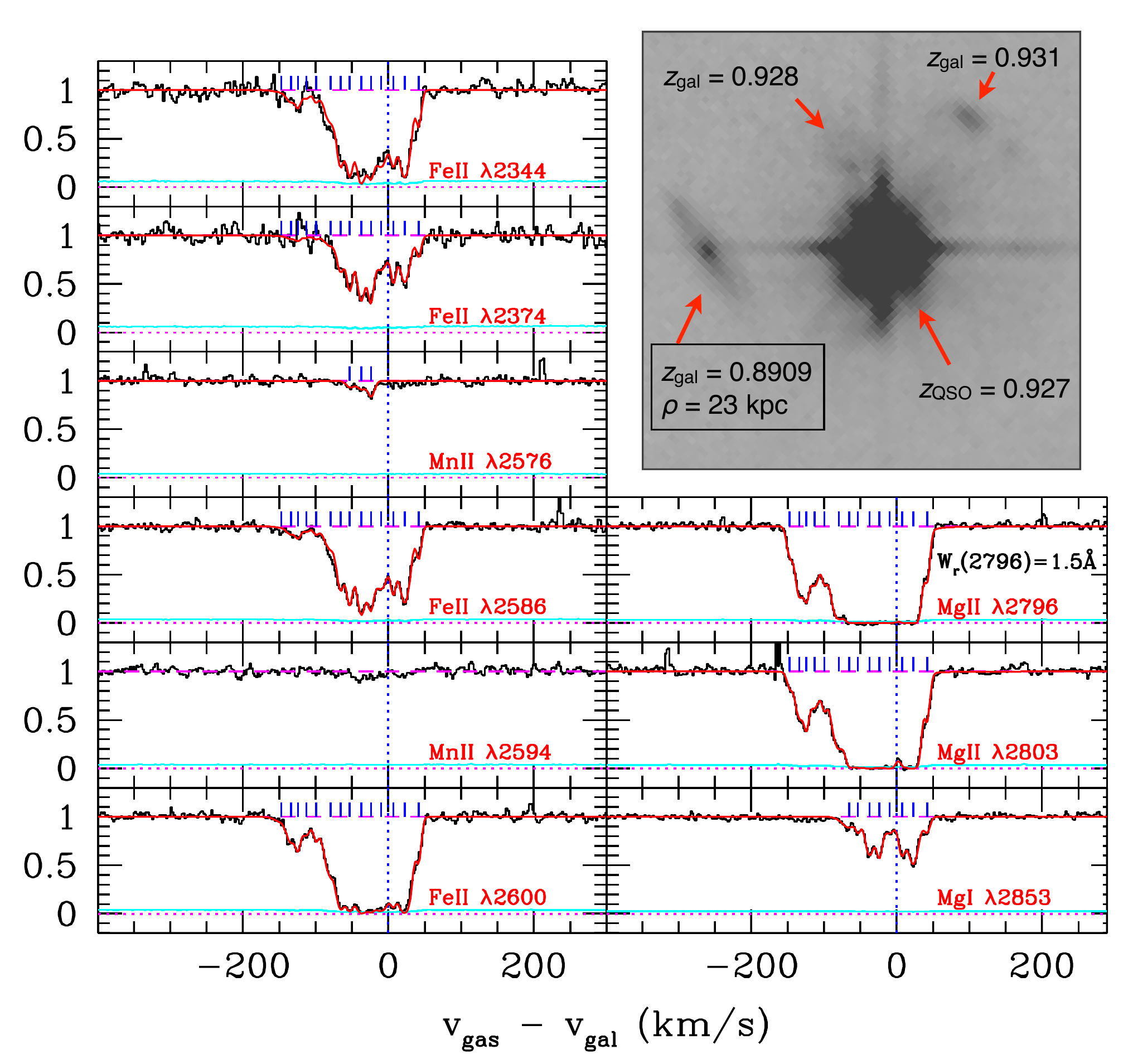}
}
\caption{Line-of-sight velocity distribution of absorbing clouds at
  projected distance $\rho=23$ kpc of Galaxy B at $z_{\rm gal}=0.891$
  with $\alpha=124.1$ degrees and $i_0=81$ degrees.  We observe strong
  absorption in Fe\,II and Mg\,II transitions.  Mn\,II and Mg\,I
  absorption features are also detected at the redshift of the galaxy,
  but are weak.  Zero velocity in each spectral panel corresponds to
  the systemic redshift of the galaxy at $z_{\rm gal}=0.8909$.  A
  Voigt profile analysis of all the observed absorption features
  together yields a minimum of 14 individual absorption components and
  $\chi^2_r=1.1$.  The total rest-frame Mg\,II absorption equivalent
  width over all observed components is $\ewr=1.5$ \AA.  The absorbing
  clumps display relative line-of-sight motions ranging from
  $\Delta\,v_{\rm los}=+42.2$ \kms\ to $\Delta\,v_{\rm los}=-147.2$
  \kms\ with respect to the systemic redshift of the galaxy.  The
  absorbing galaxy is located to the left of the QSO in the image
  panel.}
\label{3C336_z08907}
\end{figure}

For a given $\theta_0$, we determine the $z$-height at which the QSO
sightline enters ($z_1$) and exits ($z_2$) the conical outflows. If we
further assume that the line-of-sight velocity increases smoothly from
$z_1$ to $z_2$, we can calculate the appropriate range of de-projected
velocities ($v_1$, $v_2$) probed by the QSO sightline following the
framework outlined in \S\ 3.  The results are presented in Figure 3.
Our calculations show that the outflows would be accelerating if
$\theta_0\apl 10$ degrees.  Beyond $\theta_0\approx 10$ degrees, an
absorbing clump at $z_1$ would have to move faster than an absorbing
clump at $z_2$ in order to produce the observed line-of-sight velocity
of $\Delta\,v_{\rm los}^1=-54.8$ \kms, in which case the outflows
would be decelerating as the gas moves further away from the
star-forming disk.  It is straightforward to show that if we assume
decreasing line-of-sight velocity from $z_1$ to $z_2$, then the
outflows can only be decelerating over the full range of allowed
$\theta_0$.

\begin{figure}
\centerline{
\includegraphics[angle=0,scale=0.50]{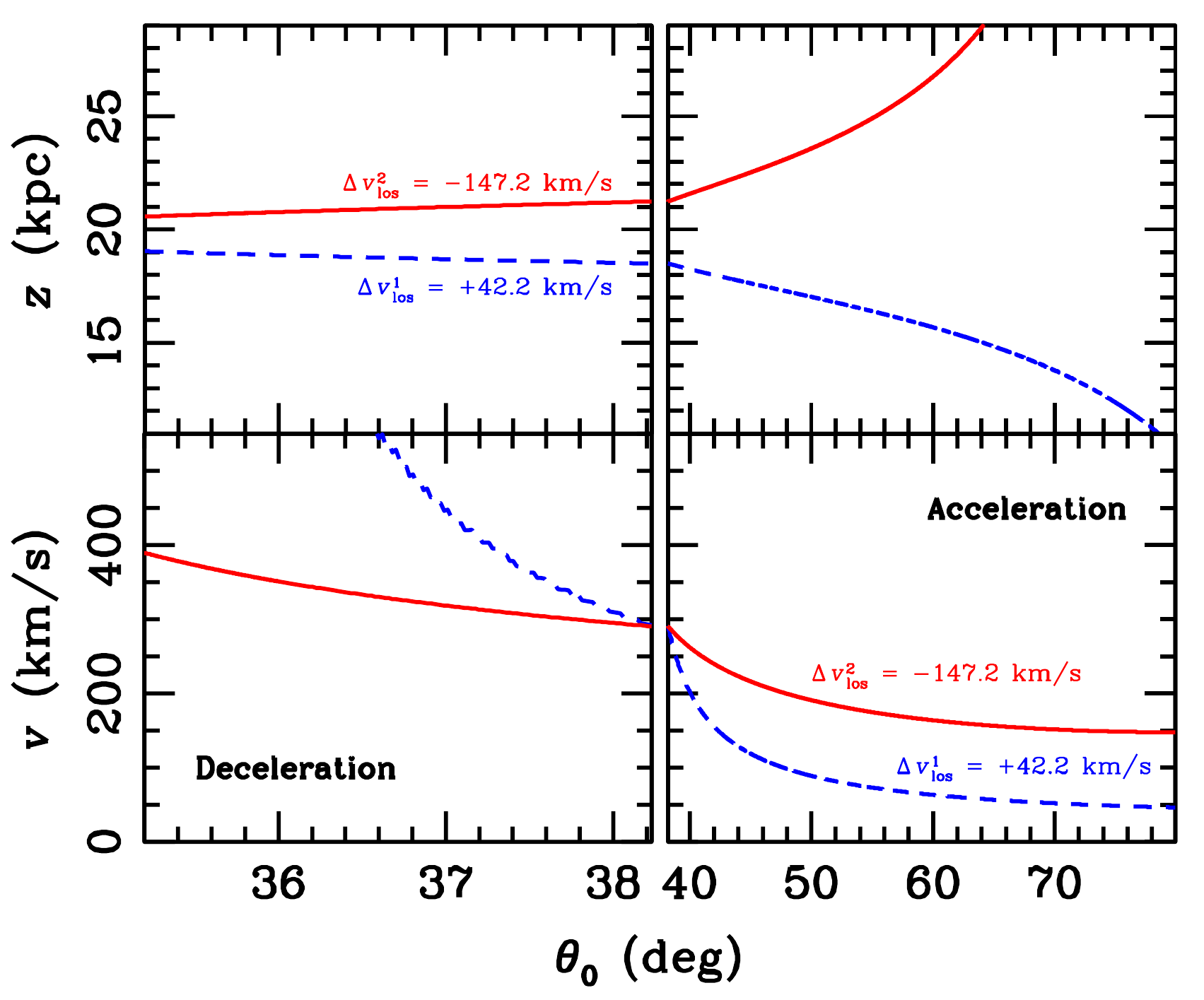}
}
\caption{Allowed parameter space for the $z$-heights (top panels) and
  de-projected velocities (bottom) of individual absorbing components
  observed in Figure \ref{3C336_z08907} versus allowed opening angle
  $\theta_0$.  The minimum allowed $\theta_0$ is constrained by the
  relative orientation of the star-forming disk with respect to the
  QSO sightline.  As shown in Figure \ref{3C336_z08907}, the galaxy is
  oriented at a position angle of $\alpha=124.1$ degrees from the QSO
  line of sight.  In order for outflows to be responsible for the
  observed absorption features in the QSO spectrum, the minimum
  allowed opening angle is $\theta_0 \apg 35$ degrees.  The presence
  of both blueshifted and redshifted components along the QSO
  sightline provides little constraint for the maximum opening angle
  of the outflow.  We place a limit at $\theta_0\approx 78$ degrees,
  beyond which the QSO sightline would be completely enclosed within
  the outflows.  Assuming an increasing line-of-sight velocity from
  $z_1$ to $z_2$ leads to deceleration for $\theta_0\apl 38.2$ degrees
  and acceleration for larger $\theta_0$.}
\label{thvza_1_z08907}
\end{figure}

\subsection{Galaxy B at $z=0.891$ in the field around 3C~336}

Galaxy B in the field aroud 3C~336 ($z_{\rm QSO} =0.927$) was
spectroscopically identifed at $z_{\rm gal}=0.8909\pm 0.0002$ by
\citet{steidel1997a}.  The galaxy is at projected distance $\rho=23.3$
kpc from the QSO line of sight.  Chen \etal\ (1998, 2001b) analyzed
available HST WFPC2 images of the field (top-right panel of Figure 4)
and measured $\alpha=124.1$ degrees and $i_0=81$ degrees for the disk
(Table 1).  The echelle spectra of the QSO cover a wavelength range
that allows observations of Fe\,II, Mn\,II, Mg\,II, and Mg\,I
absorption at the redshift of the galaxy.  The absorption profiles are
shown in individual spectral panels of Figure 4.  We detect strong
absorption complex in Fe\,II and Mg\,II transitions.  Mn\,II and Mg\,I
absorption features are also detected, but are weak.  A Voigt profile
analysis that takes into account all the observed absorption
transitions yields a minimum of 14 individual absorption components
and $\chi^2_r=1.1$.  The total rest-frame Mg\,II absorption equivalent
width over all observed components is $\ewr=1.53\pm 0.05$ \AA.  The
absorbing clumps display relative line-of-sight motions ranging from
$\Delta\,v_{\rm los}=+42.2$ \kms\ to $\Delta\,v_{\rm los}=-147.2$
\kms\ with respect to the systemic redshift of the galaxy.

\begin{figure}
\centerline{
\includegraphics[angle=0,scale=0.40]{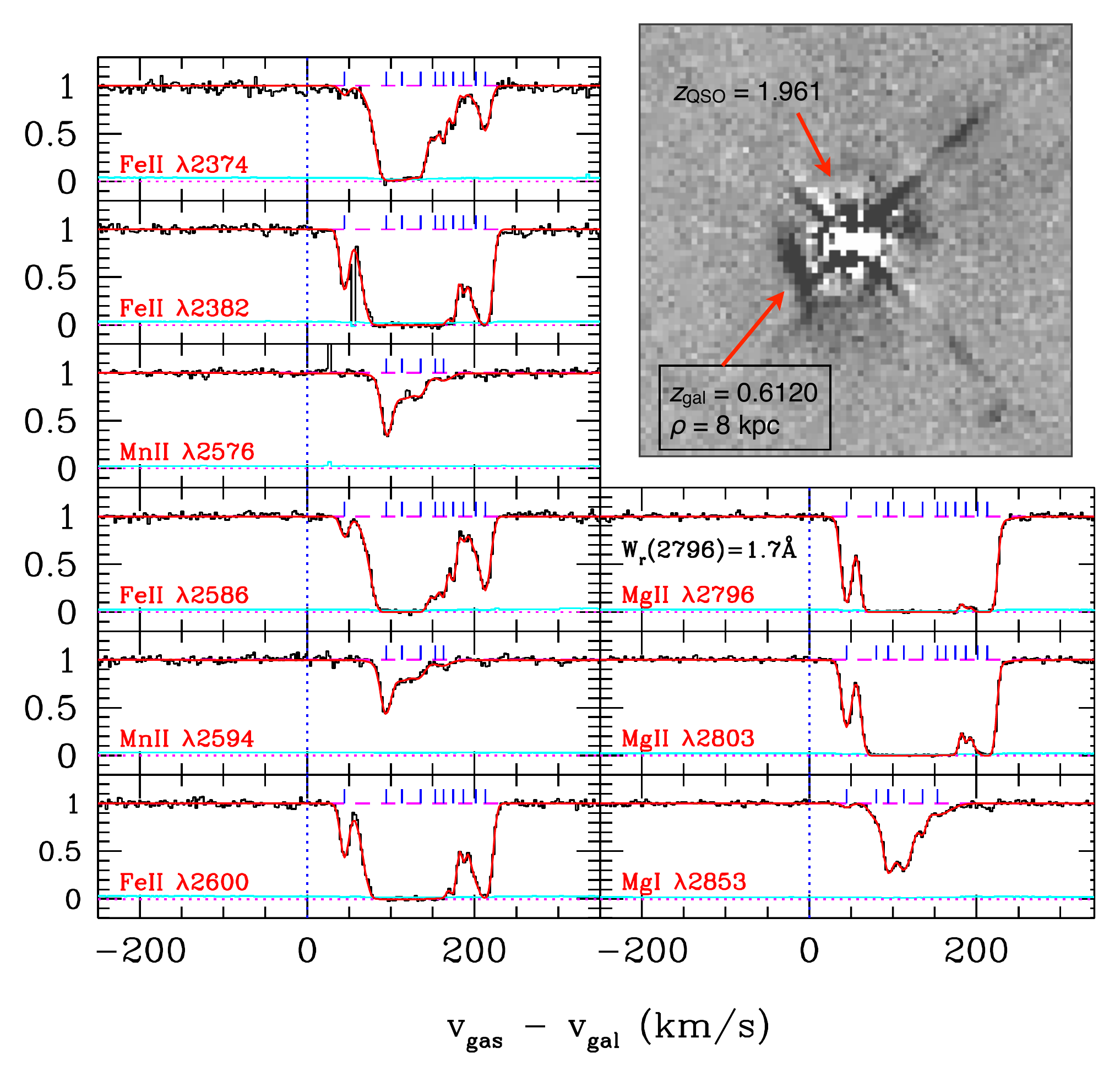}
}
\caption{Line-of-sight velocity distribution of absorbing clouds at
  projected distance $\rho=8$ kpc of Galaxy C at $z_{\rm gal}=0.612$
  with $\alpha=113$ degrees and $i_0=65$ degrees.  We observe strong
  absorption in Fe\,II, Mn\,II, Mg\,II, and Mg\,I transitions at the
  redshift of the galaxy.  Zero velocity in each spectral panel
  corresponds to the systemic redshift of the galaxy at $z_{\rm
    gal}=0.6120$.  A Voigt profile analysis of the observed Fe\,II,
  Mn\,II, Mg\,II, and Mg\,I absorption profiles yields a minimum of 11
  individual absorption components and $\chi^2_r=1.1$.  The total
  rest-frame Mg\,II absorption equivalent width over all observed
  components is $\ewr=1.7$ \AA.  The absorbing clumps display relative
  line-of-sight motions ranging from $\Delta\,v_{\rm los}=+44.9$ \kms\
  to $\Delta\,v_{\rm los}=+213.3$ \kms\ with respect to the systemic
  redshift of the galaxy.  The absorbing galaxy is blended with the
  QSO light.  The upper right panel displays the galaxy after removing
  the point spread function of the QSO.}
\label{Q0058_z06120}
\end{figure}

Figure 4 shows that the cool gas probed by the Mg\,II absorption
transitions exhibits both blueshifted and redshifted motion with
respect to the star-forming disk with a total line-of-sight velocity
spread of $\approx 190$\kms.  Following the discussion for Galaxy A in
\S\ 4.1, the position angle of the galaxy $\alpha=124.1$ degrees from
the QSO line of sight requires that the opening angle be greater than
$\theta_0\approx 35$ degrees in order for outflows to be responsible
for the observed absorption features in the QSO spectrum.  However,
the presence of both blueshifted and redshifted components along the
QSO sightline provides little constraint for the maximum opening angle
of the outflow.  We place a limit at $\theta_0\approx 78$ degrees,
beyond which because the inferred $z$-height blows up to unrealistic
values.

Following the discussion in \S\ 4.1, we can calculate the allowed
values for the $z$-height and de-projected velocity of the outflows
from Galaxy B by assuming that the line-of-sight velocity increases
smoothly from $z_1$ to $z_2$.  The results are presented in Figure 5.
Different from Galaxy A, our calculations show that the outflows from
Galaxy B would be decelerating if $\theta_0\apl 38.2$ degrees.  Beyond
$\theta_0\approx 38.2$ degrees, an absorbing clump at $z_1$ would be
moving at a lower velocity than those at $z_2$ in order to produce the
observed line-of-sight velocity of $\Delta\,v_{\rm los}^1=+42.2$ \kms.
In this case, the outflows would be accelerating as the gas moves
further away from the star-forming disk for a broad range of
$\theta_0$.  Similar to Galaxy A, it is straightforward to show that
if we assume decreasing line-of-sight velocity from $z_1$ to $z_2$,
then the outflows can only be decelerating over the full range of
allowed $\theta_0$.

\begin{figure}
\centerline{
\includegraphics[angle=0,scale=0.50]{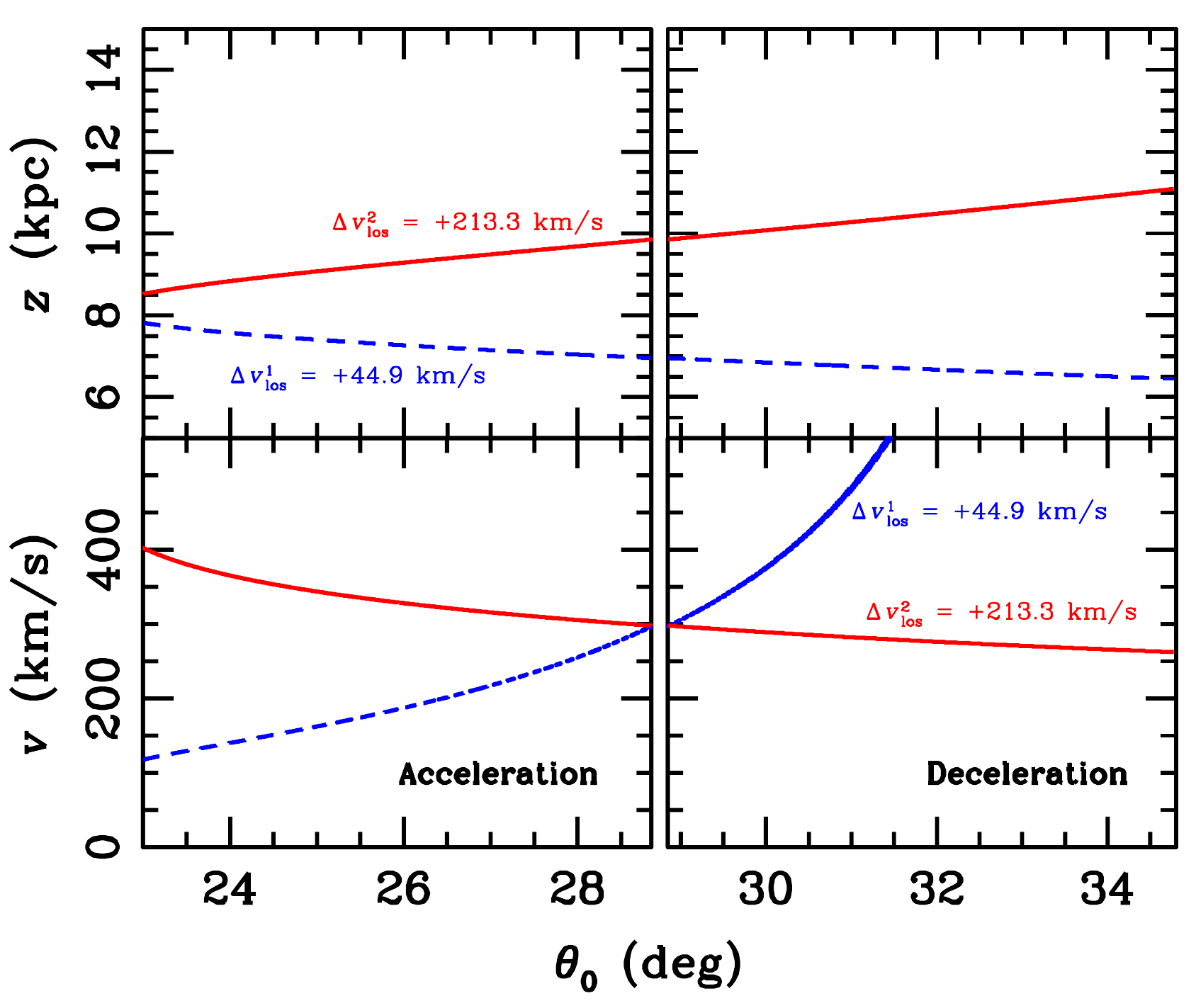}
}
\caption{Allowed parameter space for the $z$-heights (top panels) and
  de-projected velocities (bottom) of individual absorbing components
  observed in Figure 6 versus allowed opening angle $\theta_0$.  The
  minimum and maximum allowed $\theta_0$ are constrained by the
  relative orientation and alignment of the star-forming disk with
  respect to the QSO sightline.  As shown in Figure 6, the galaxy is
  oriented at a position angle of $\alpha=113$ degrees from the QSO
  line of sight.  In order for outflows to be responsible for the
  observed absorption features in the QSO spectrum, the minimum
  allowed opening angle is $\theta_0 \apg 23$ degrees.  In addition,
  all absorbing clumps are found redshifted from the systemic redshift
  of the galaxy.  Given that the galaxy has a inclination angle of
  $i_0=65$ degrees, the lack of blueshifted absorbing components
  constrains the opening angle at $\theta_0\apl 35$ degrees.  Assuming
  an increasing line-of-sight velocity from $z_1$ to $z_2$ leads to
  acceleration for $\theta_0\apl 29$ degrees and deceleration for
  larger $\theta_0$.}
\label{thvza_1_q0058}
\end{figure}

\subsection{Galaxy C at $z=0.612$ in the field around LBQS~$0058+0155$}

Galaxy C in the field aroud LBQS~$0058+0155$ ($z_{\rm QSO} =1.954 $)
was spectroscopically identifed at $z_{\rm gal}=0.6120\pm 0.0002$ by
Chen \etal\ (2005).  The galaxy is at projected distance $\rho=7.9$
kpc from the QSO line of sight.  Pettini \etal\ (2000) analyzed
available HST WFPC2 images of the field (top-right panel of Figure 6)
and estimated $i_0\approx 65$ degrees for the disk (Table 1).  We
analyzed the images ourselves and estimated $\alpha=113$ degrees.  The
echelle spectra of the QSO cover a wavelength range that allows
observations of Fe\,II, Mn\,II, Mg\,II, and Mg\,I absorption at the
redshift of the galaxy.  The absorption profiles are shown in
individual spectral panels of Figure 6.  We detect a strong absorption
complex in Fe\,II, Mn\,II, Mg\,II, and Mg\,I transitions.  A Voigt
profile analysis that takes into account all the observed absorption
profiles yields a minimum of 11 individual absorption components and
$\chi^2_r=1.1$.  The total rest-frame Mg\,II absorption equivalent
width over all observed components is $\ewr=1.67\pm 0.01$ \AA.  The
absorbing clumps display relative line-of-sight motions ranging from
$\Delta\,v_{\rm los}=+44.9$ \kms\ to $\Delta\,v_{\rm los}=+213.3$
\kms\ with respect to the systemic redshift of the galaxy.

Figure 6 shows that the cool gas probed by the Mg\,II absorption
transitions is entirely redshifted with respect to the star-forming
disk with a total line-of-sight velocity spread of $\approx 170$\kms.
Following the discussion in \S\ 4.1, the position angle of the galaxy
$\alpha=113$ degrees from the QSO line of sight requires that the
opening angle be greater than $\theta_0\approx 23$ degrees in order
for outflows to be responsible for the observed absorption features in
the QSO spectrum.  In addition, the galaxy has a inclination angle of
$i_0=65$ degrees, and therefore the lack of blueshifted absorbing
components constrains the opening angle at $\theta_0\apl 35$ degrees.

Following the discussion in \S\ 4.1, we can calculate the allowed
values for the $z$-height and de-projected velocity of the outflows
from Galaxy C by assuming that the line-of-sight velocity increases
smoothly from $z_1$ to $z_2$.  The results are presented in Figure 7.
Similar to Galaxy A, our calculations show that the outflows from
Galaxy C would be accelerating if $\theta_0\apl 29$ degrees.  Beyond
$\theta_0\approx 29$ degrees, an absorbing clump at $z_1$ would be
moving at a larger velocity than those at $z_2$ in order to produce
the observed line-of-sight velocity of $\Delta\,v_{\rm los}^1=+44.9$
\kms.  In this case, the outflows would be decelerating as the gas
moves further away from the star-forming disk.  Similar to Galaxies A
and B, it is straightforward to show that if we assume decreasing
line-of-sight velocity from $z_1$ to $z_2$, then the outflows can only
be decelerating over the full range of allowed $\theta_0$.  

We note that the absorber is a known damped \lya\ absorption system at
$z=0.612$ with neutral hydrogen column density of $\log\,N({\rm
  H\,I})=20.1\pm 0.2$ (Pettini \etal\ 2000).  It is likely that a
significant fraction of the observed absorption originates in the
star-forming disk.

\begin{figure}
\centerline{
\includegraphics[angle=0,scale=0.40]{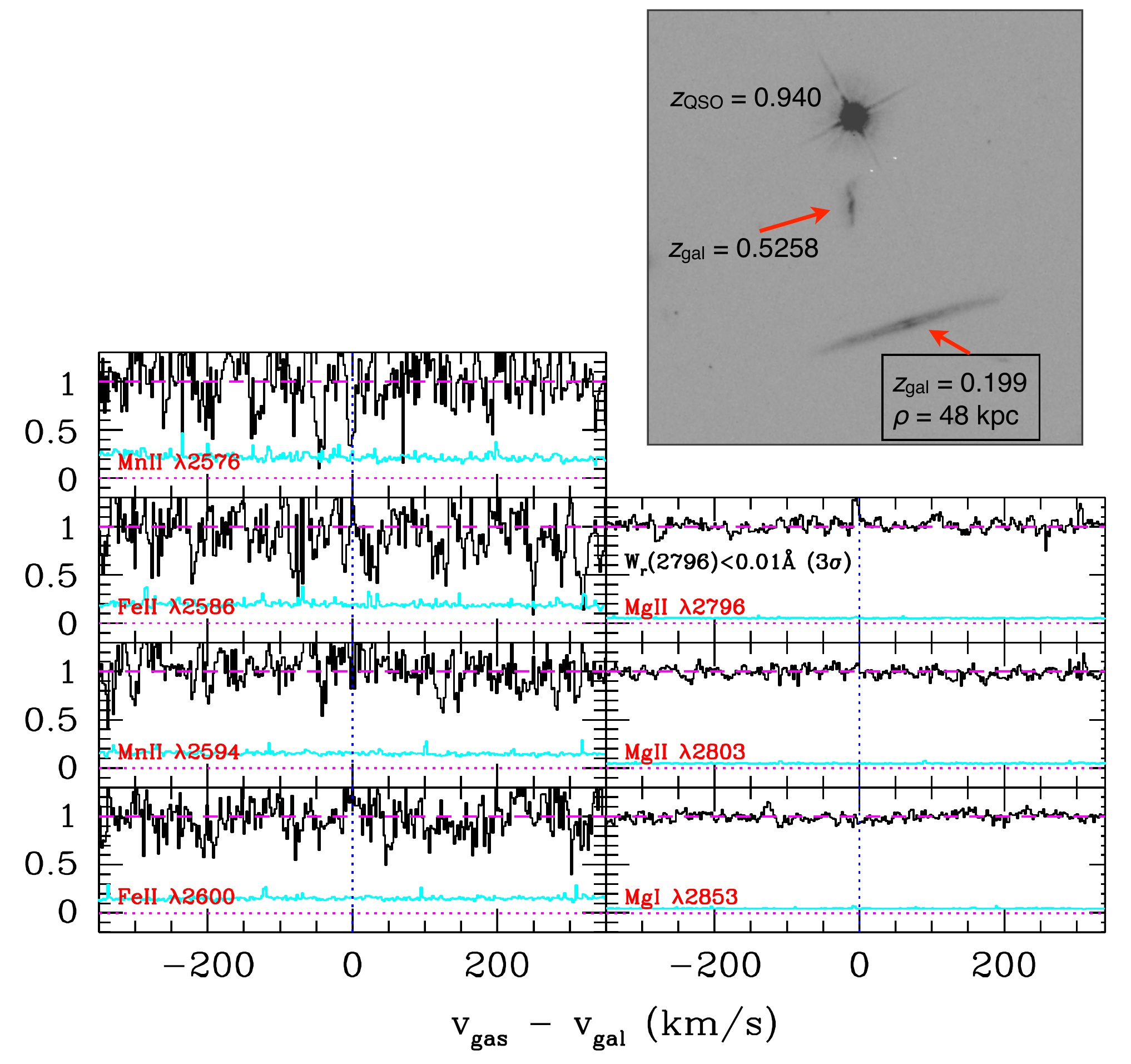}
}
\caption{Absorption properties of Galaxy D at $z_{\rm gal}=0.199$ in
  the field around QSO Q0827$+$243.  The image in the upper right
  panel shows a nearly edge-on galaxy at $\rho=48$ kpc.  No absorption
  feature is detected at the redshift of the galaxy in the QSO
  spectrum.  We measure a 3-$\sigma$ upper limit to the Mg\,II
  absorption strength of $W_r(2796)=0.01$ \AA.}
\label{Q0827}
\end{figure}

\subsection{Galaxy D at $z=0.199$  in the field toward Q$0827+243$}

Galaxy D in the field aroud Q~$0827+243$ ($z_{\rm QSO} =0.939$) was
spectroscopically identifed at $z_{\rm gal}=0.199$ by Steidel \etal\
(2002).  The nearly edge-on galaxy is at projected distance $\rho=48$
kpc from the QSO line of sight.  We analyzed available HST images of
the field and estimated $i_0=85$ degrees and $\alpha=86.3$ degrees.
The echelle spectra of the QSO cover a wavelength range that allows
observations of Fe\,II, Mn\,II, Mg\,II, and Mg\,I absorption at the
redshift of the galaxy.  However, we detect no trace of absorption
features at the redshift of the galaxy in the QSO spectrum (Figure 8).
We determine a 3-$\sigma$ upper limit to the Mg\,II absorption
strength of $W_r(2796)=0.01$ \AA.  The lack of absorption features
associated with Galaxy D at $z_{\rm gal}=0.199$ implies that
large-scale galactic outflows may not exist in this galaxy, despite a
clear dust lane feature along the disk in the high-resolution HST
image (upper right panel of Figure 8).  Alternatively, the opening
angle of the outflows may be small, $\theta_0\apl 4$ degrees, or
outflows from the star-forming disk do not reach out to $\sim 50$ kpc.

A second edge-on galaxy is seen at $\rho=38$ kpc and $z_{\rm
  gal}=0.526$.  This galaxy has been studied in detail by Steidel
\etal\ (2002) and Chen \etal\ (2005).  The major axis of this galaxy
is oriented directly toward the QSO line of sight with $\alpha=0$.  We
therefore exclude the galaxy from the study presented here.

\begin{footnotesize}
\begin{table*}
\centering
\begin{minipage}{120mm}
\caption{Allowed Parameter Space for Accelerated Outflows}
\begin{tabular}{ccccccc}
\hline
 & $\Delta\,v_{\,\rm los}$ & \multicolumn{1}{c}{$\theta_{\rm 0}$} & $z_1$ & $z_2$ & $v_1$ & $v_2$ \\
 Galaxy & (km/s) & \multicolumn{1}{c}{$(^{\circ})$} & (kpc) & (kpc) & (km/s) & (km/s) \\
\hline
\hline
A   & [-54.8, -143.7] &  [6.5, 9.8] & [34.7, 33.5] & [34.7, 36.1] & [200, 358] & [524, 361]\\
B   & [+42.2, -147.2] & [38.2, 78.0] & [18.5, 10.3] & [20.2, 240.2] & [286, 46] & [289, 147] \\ %
C   & [+44.9, +213.3] & [23.0, 28.9] &   [7.8, 7.0] & [8.5, 9.9] & [118, 297] & [402, 297]  \\ %
\hline
\label{model_results_a}
\end{tabular}
\end{minipage}
\end{table*}
\end{footnotesize}
 
\section{Discussion}

Under the hypothesis that Mg\,II absorbers found near the minor axis
of a disk galaxy originate in the cool phase of super-galactic winds,
we have carried out a study to constrain the properties of large-scale
galactic outflows at redshift $z_{\rm gal}\apg 0.5$ based on the
observed relative motions of individual absorbing clouds with respect
to the positions and orientations of the absorbing galaxies.  We have
identified in the literature four highly inclined disk galaxies
located within 50 kpc and with the minor axis oriented within 45
degrees of a background QSO sightline. Deep HST images of the galaxies
are available for accurate characterizations of the optical
morphologies of the galaxies.  High-quality echelle spectra of the QSO
members are also available in public archives for resolving the
velocity field of individual absorption clumps.  All but one of the
four galaxies in our study exhibit a strong associated Mg\,II
absorption feature with $\ewr \apg 0.8$ \AA\ at $\rho=8-34$ kpc.  If
super-galactic winds are present in all star-forming galaxies, then
the absence of Mg\,II absorber to a 3-$\sigma$ upper limit of
$\ewr=0.01$ \AA\ at $\rho=48$ kpc around the non-absorbing galaxy (D)
indicates that either the opening angle of the outflows is small,
$\theta_0\apl 4$ degrees, or outflows from the star-forming disk do
not reach out to $\sim 50$ kpc.

Combining known morphological parameters of the galaxies such as the
inclination ($i_0$) and orientation ($\alpha$) angles of the
star-forming disks, and resolved absorption profiles of the associated
absorbers at $< 35$ kpc away, we have explored the allowed parameter
space for the opening angle $\theta_0$ and the velocity field of
large-scale galactic outflows as a function of $z$-height, $v(z)$,
from each disk galaxy.  In this section, we discuss the implications
of our analysis.

\subsection{Kinematics of Super-galactic Winds}

The results of our analysis presented in \S\ 4 show that the observed
absorption signatures of the Mg\,II doublets and the Fe\,II series are
compatible with the absorbing gas being either accelerated or
decelerated (Figures 3, 5, \& 7) for different ranges of the opening
angle of the outflows.  We summarize the allowed parameter space for
accelerated outflows in Table 2, which lists for each galaxy the
observed range of relative line-of-sight velocities of individual
absorbing components with respect to the systemic velocity of the
galaxy ($\Delta\,v_{\rm los}$), the range of opening angle
($\theta_0$) over which the observed velocity spread can by explain by
accelerated outflows, the corresponding range of $z$-height above the
star-forming disk ($z_1$) where the QSO sightline enters the
accelerated outflows (Figure 1), the corresponding range of $z_2$
where the QSO sightline exits the outflows, and the corresponding
ranges of outflowing velocities at $z_1$ ($v_1$) and $z_2$ ($v_2$).

As described in \S\ 4, our calculations are based on the assumption
that the outflow velocities follow a smooth gradient with the distance
from the star-forming disk, which allows us to project the observe
velocity components to the appropriate $z$-heights in the conical
outflow model.  Assuming that the observed line-of-sight velocity
increases with the $z$-height, Table 2 shows that acceleration is
valid only for a limited range of $\theta_0$ for galaxies A and C.
Beyond the maximum allowed angle in each of the galaxies, the
outflowing gas at $z_1$ would have to move faster than those at larger
$z$-heights in order to reproduce the small projected velocity found
along the QSO sightline.  The outflows would be decelerating, instead
of accelerating.  For galaxy B, the upper bound for $\theta_0$ remains
unconstrained, but the inferred $z$-height blows up to unphysically
large values at $\theta_0=78$ degrees.

\begin{figure}
\centerline{
\includegraphics[angle=0,scale=0.450]{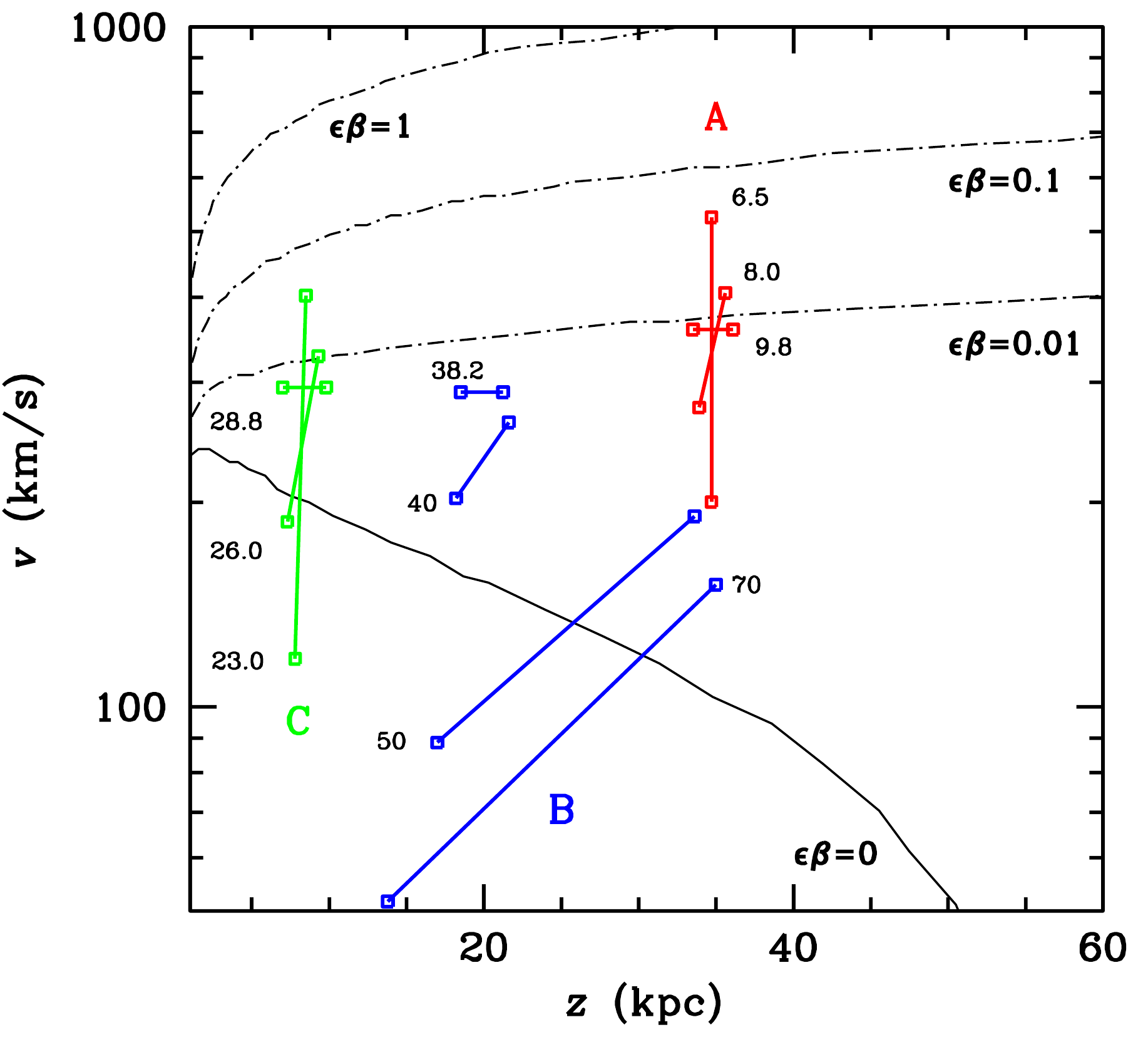}
}
\caption{Comparisons between the deprojected outflow velocities as a
  function of $z$-height $v(z)$ observed around star-forming disks and
  model predictions for accelerated outflows.  For each galaxy, we
  present the velocity range and the corresponding $z$-height in the
  large-scale outflows for a set of $\theta_0$ that cover the full
  range of allowed $\theta_0$ as summarized in Table 2.  Each set of
  data points is labled by the corresponding $\theta_0$.  It is clear
  that for a significant fraction of allowed $\theta_0$ the inferred
  velocity gradient is extreme, increasing outflow velocity by more
  than 100\% for only a few percent gain in $z$-height.  We adopt
  model predictions from \citet{murray2011a}, who calculated the
  velocity field of super-galactic winds driven by the radiation and
  ram pressure forces from a stellar cluster of $10^6\,{\rm M}_\odot$
  in an M82-like galaxy.  The dot-dashed curves are for different
  feedback efficiencies $\epsilon\,\beta$, where $\epsilon$ is the
  fraction of the supernova luminosity that is thermalized to produce
  a hot phase and $\beta$ is the mass loading factor limited within
  the range, $1\le \beta\le 17$.  In the absence of a ram pressure
  drag force, the predicted velocity field is shown in the solid curve
  with $\epsilon\,\beta=0$.  We find that for every galaxy the Murray
  et al.\ model can explain the observations for only a narrow range
  of $\theta_0$ where the inferred acceleration is minimal.  Namely,
  $\theta_0\approx 10$ degrees for galaxy A, $\theta_0\approx 38$
  degrees for galaxy B, and $\theta_0\approx 29$ degrees for galaxy C.
  At these specified opening angles, the feedback efficiency is small
  with $\epsilon\,\beta\apl 0.01$.}
\label{mod_comp_acc}
\end{figure}

We present in Figure 9 the observed velocity field versus $z$-height
above the star-forming disk of each galaxy for a set of $\theta_0$
that cover the full range of allowed $\theta_0$ as summarized in Table
2.  It is clear that for a significant fraction of allowed $\theta_0$,
the inferred velocity gradient is extreme.  A few percent gain in
$z$-height would result in increasing outflow velocity by more than
100\%, further narrowing down the range of reasonable $\theta_0$ that
would yield a more physical acceleration field.

For comparison, we consider model predictions by Murray \etal\ (2011;
hereafter M11), who presented an analytical model for launching
large-scale galactic winds.  These authors showed that radiation
pressure from the most massive star clusters in the disk can clear
holes in the disk, allowing subsequent supernovae ejecta to escape the
disk.  As the outflowing material is lifted above the disk, the
combined influence of radiation and ram pressure forces from multiple
star clusters would then accelerate the winds further out to several
tens of kpc away from the disk.

M11 provided model predictions for different types of galaxies,
including M82, the Milky Way and luminous starburst galaxies at
$z\sim2$.  For each galaxy type, the authors considered the effect of
different physical parameters on the kinematics of outflowing gas,
including the ram pressure drag force due to the hot supernova winds
and the mass of super star clusters.  Given that the galaxies in our
sample are sub-$L_*$ galaxies (Table 1), we will focus our comparisons
on the M82-like models.

Specifically, we adopt the predicted velocity field from M11 for
large-scale galactic outflows that are driven by the combined
influence of radiation and ram pressure forces of a super star cluster
of $10^6\,{\rm M}_\odot$ in an M82-like galaxy.  The dot-dashed curves
in Figure 9 indicate the predictions for different feedback
efficiencies $\epsilon\,\beta$, where $\epsilon$ is the fraction of
the supernova luminosity that is thermalized to produce a hot phase
and $\beta$ is the mass loading factor limited within the range, $1\le
\beta\le 17$.  In the absence of a ram pressure drag force, the
predicted velocity field is represented by the solid curve in Figure 9
with $\epsilon\,\beta=0$.

Figure 9 shows that in order for the M11 model to be consistent with
observations, the allowed $\theta_0$ are further reduced to a narrow
range, where the inferred acceleration is minimal.  Namely,
$\theta_0\approx 10$ degrees for galaxy A, $\theta_0\approx 38$
degrees for galaxy B, and $\theta_0\approx 29$ degrees for galaxy C.
At these specified $\theta_0$, the observations also favor a small
feedback efficiency with $\epsilon\,\beta\apl 0.01$.

\begin{figure}
\centerline{
\includegraphics[angle=0,scale=0.450]{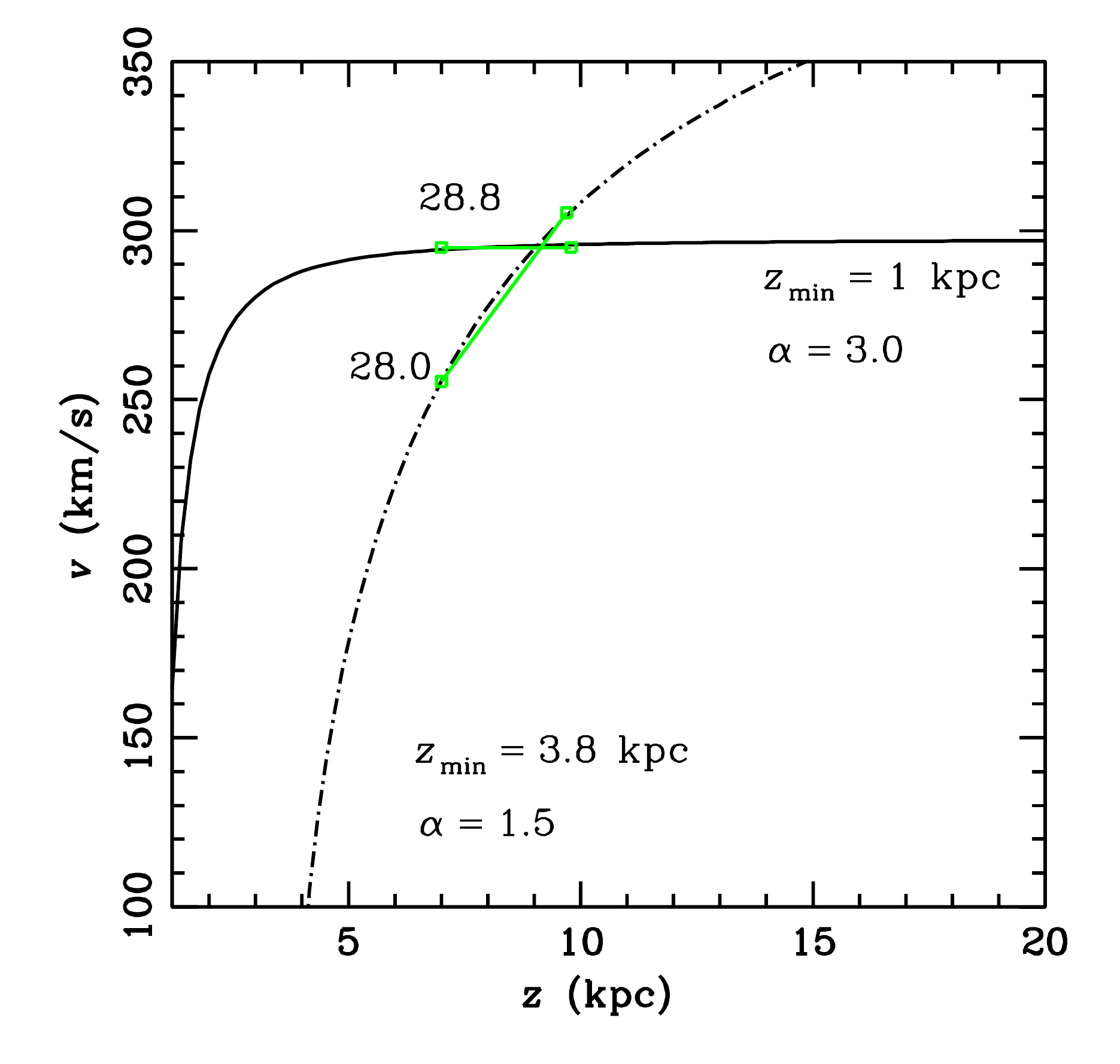}
}
\caption{Examples to illustrate the appropriate power-law model for
  characterizing accelerated outflows inferred for our sample
  galaxies.  We consider two specific cases for galaxy C at $\rho=7.9$
  kpc.  As summarized in Table 2, the allowed opening angle of
  accelerated outflows in this galaxy is $\theta_0=23-28.9$ degrees.
  The plot shows the derived $v(z)$ for $\theta_0=28.8$ degrees and
  $\theta_0=28$ degrees.  In order to reproduce the constant velocity
  field at $z\approx 10$ kpc for an opening angle of $\theta_0=28.8$
  degrees, the acceleration model formulated in Equations (6) and (7)
  should have a steep power-law index of $\alpha=3$ for a launch
  radius of $z_{\rm min}=1$ kpc.  To reproduce the large velocity
  gradient implied by a slightly smaller opening angle $\theta_0=28$
  degrees requires a launch radius of $z_{\rm min}=3.8$ kpc for a
  shallower power-law index $\alpha=1.5$ that is more typical of what
  is found in $z_{\rm gal}=2-3$ starburst galaxies (Steidel \etal\
  2010).  For smaller $\theta_0$, the velocity gradient steepens
  (Figure 9).  While the shallower power-law index is a viable
  solution, the launch radius would approach $z_{\rm min}\approx 7.5$
  kpc.  By analogy, we conclude that adopting a shallower power-law
  index $\alpha\apl 2$ for characterizing the accelerated outflows in
  galaxies A \& B (Figure 9) would require a launch radius $z_{\rm
    min} \apg 20$ kpc.}
\label{mod_comp_acc}
\end{figure}

The narrow range of favored $\theta_0$ from the model comparisons in
Figure 9 is understood by the expected decline of radiation and ram
pressure forces at distances beyond 1 kpc (Figure 2 in M11).  To
characterize the declining acceleration as a function of $z$-height,
we consider a power-law model following the parameterization of
Steidel \etal\ (2010) (see also Veilleux \etal\ (1994) who adopted 
a similar parametrization),
\begin{equation}
a(z) = A\,z^{-\alpha}.
\end{equation}
If we designate $z_{\rm min}$ as the launch $z$-height of
super-galactic winds, then we can recast Equation (6) in terms of the
outflows velocity field as
\begin{equation}
v(z) = \left(\frac{2\,A}{\alpha-1}\right)^{1/2}\,\sqrt{z_{\rm min}^{1-\alpha}-z^{1-\alpha}}.
\end{equation}
Steidel \etal\ (2010) found that the acceleration model defined in
Equations (6) and (7) with $\alpha=1.15-1.95$ and $z_{\rm min}=1$ kpc
can reproduce the blueshifted self-absorption of low-ionization
transitions found in luminous starburst galaxies at $z_{\rm gal}=2-3$.

To explore the power-law index $\alpha$ and launch radius $z_{\rm
  min}$ appropriate for describing the accelerated outflows in our
sample in Figure 9, we consider two specific cases for galaxy C at
$\rho=7.9$ kpc.  As summarized in Table 2, the allowed opening angle
of accelerated outflows in this galaxy is $\theta_0=23-28.9$ degrees.
Figure 10 displays the derived $v(z)$ for $\theta_0=28.8$ and
$\theta_0=28$ degrees.  At $\theta_0=28.8$ degrees, the conical
outflows model expects that the QSO sightline probes outflowing gas
from $z_1=7.9$ kpc to $z_2=11.2$ kpc with little change in the outflow
velocity.  At $\theta_0=28$ degrees, the conical outflows model
expects that the QSO sightline probes outflowing gas from $z_1=7.9$
kpc to $z_2=11$ kpc with the outflow velocity increasing from
$v_1=255$ \kms\ to $v_2=305$ \kms.

\begin{footnotesize}
\begin{table*}
\centering
\begin{minipage}{120mm}
\caption{Allowed Parameter Space for Decelerated Outflows}
\begin{tabular}{ccccccc}
\hline
 & $\Delta\,v_{\,\rm los}$ & \multicolumn{1}{c}{$\theta_{\rm 0}$} & $z_1$ & $z_2$ & $v_1$ & $v_2$ \\
 Galaxy & (km/s) & \multicolumn{1}{c}{$(^{\circ})$} & (kpc) & (kpc) & (km/s) & (km/s) \\
\hline
\hline
A   & [-54.8, -143.7] & [9.8, 17.3] & [33.5, 32.1] & [36.1, 37.9] & [361, 5085] & [361, 270] \\
    & [-143.7, -54.8] & [6.5, 17.3] & [34.7, 32.1] & [34.7, 37.9] & [525, 13335] & [200, 103] \\
B   & [+42.2, -147.2] & [35.2, 38.2] & [19.0, 18.5] & [20.6, 21.2] & [4320, 291] & [390, 290] \\ %
    & [-147.2, +42.2] & [35.2, 78.0] & [19.0, 10.3] & [20.6, 240.2] & [15070, 160] & [112, 42] \\ %
C   & [+44.9, +213.3] & [28.9, 34.8] &   [7.0, 6.4] & [9.9, 11.1] & [297, 7708] & [298, 262]  \\ %
    & [+213.3, +44.9] & [23.0, 34.8] &   [7.8, 6.4] & [8.5, 11.1] & [559, 36617] & [85, 55]  \\ %
\hline
\label{model_results_a}
\end{tabular}
\end{minipage}
\end{table*}
\end{footnotesize}
 
Following the acceleration model formulated in Equations (6) and (7),
to reproduce the constant velocity field at $z\approx 10$ kpc for an
opening angle of $\theta_0=28.8$ degrees requires a steep power-law
index of $\alpha=3$ for a launch radius of $z_{\rm min}=1$ kpc.  To
reproduce the large velocity gradient implied by a slightly smaller
opening angle $\theta_0=28$ degrees requires a launch radius of
$z_{\rm min}=3.8$ kpc for a shallower power-law index $\alpha=1.5$
that is more typical of what is found in $z_{\rm gal}=2-3$ starburst
galaxies (Steidel \etal\ 2010).  As summarized in Table 2, the allowed
opening angle of galaxy C for accelerated outflows can be as small as
$\theta_0=23$ degrees.  For $\theta_0=23$ degrees, the shallower
power-law index is a viable solution but the launch radius needs to
increase to $z_{\rm min}\approx 7.5$ kpc.  By analogy, we conclude
that adopting a shallower power-law index $\alpha\apl 2$ for
characterizing the accelerated outflows in galaxies A \& B would
require a launch radius $z_{\rm min} \apg 20$ kpc.

In summary, we have shown that acceleration is only possible for a
limited range of $\theta_0$, if we project the velocity components to
the appropriate $z$-height assuming that the observed line-of-sight
velocity increases with $z$-height.  Under an acceleration scenario,
the observations favor a narrow range of model parameters such as
$\epsilon\,\beta \apl 0.01$, and $\alpha\approx 3$ for $z_{\rm min}=1$
kpc or $\alpha\approx 1.5$ for $z_{\rm min}\apg 4$ kpc.  For
$\theta_0$ outside of the range summarized in Table 2, we have shown
that the gas would be decelerating, because a larger speed is required
at smaller $z$-height in order to reproduce the small projected
velocity along the QSO sightline.  In addition, if we adopt an inverse
mapping between the line-of-sight velocity and $z$-height, namely if
the observed line-of-sight velocity decreases with increasing
$z$-height, then the outflows would only be decelerating over the full
range of allowed $\theta_0$.  It is clear that the allowed parameter
space for decelerated outflows is significantly broader than what is
shown in Table 2, but Figures 3, 5, \& 7 have shown that beyond
certain $\theta_0$ the de-projected velocity in the outflows exceeds
1000 \kms\ at $z>10$ kpc which also seem unphysical.  A summary for
decelerated outflows is presented in Table 3.

\subsection{Implications for the Origin of Super-Galactic Winds}

We have shown that combining known inclination ($i_0$) and orientation
($\alpha$) angles of the star-forming disks and resolved absorption
profiles of the associated absorbers at projected distances
$\rho=8-35$ kpc, we can derive strong constraints for the opening
angle $\theta_0$, the velocity field of large-scale galactic outflows
$v(z)$ which leads to constraints for the supernova feedback
efficiency $\epsilon\,\beta\apl 0.01$, and the acceleration parameters
$\alpha$ and $z_{\rm min}$.  Here we examine whether these parameter
constraints are reasonable and discuss their implications.

First, we consider the allowed opening angle $\theta_0$ for
accelerated outflows.  Table 2 shows that with the exception of galaxy
B, for which the upper bound of $\theta_0$ remains unconstrained, both
galaxies A and C are constrained to have $\theta_0< 10$ and
$\theta_0<30$ degrees, respectively, for the outflows probed by Mg\,II
and Fe\,II absorption transitions.  These allowed values of $\theta_0$
are smaller than the typical opening angle of $2\,\theta_0 \approx
60-135$ degrees seen in nearby starburst galaxies (e.g.\ Veilleux
\etal\ 2005).  The discrepancy can be understood, if there exist a
tilt or asymmetries around the minor axis in the outflows (e.g.\
Shopbell \& Bland-Hawthorn 1998; Sugai \etal\ 2003), in which case the
opening angle of the outflows in our galaxies may be bigger.  In
addition, we note that these constraints are derived for Mg\,II
absorbing gas which presumably traces the cool phase of super-galactic
winds.  Hot galactic winds can be more wide spread than the cool
clumps.  Numerical simulations have shown that cool clouds embedded in
a supersonic wind are broken up to form elongated filamentary
structures (Cooper \etal\ 2009).  The small $\theta_0$ found in our
study may be understood, if the Mg\,II absorbers are produced in these
elongated filaments.

Next, we consider the constraint for supernova feedback efficiency
$\epsilon\,\beta\apl 0.01$.  Recall that in the Murray \etal\ (2011)
model, $\epsilon$ is the fraction of the supernova luminosity that is
thermalized to produce a hot phase and $\beta$ is the mass loading
factor which by definition must be $\beta>1$.  The constraint we found
therefore requires that the efficiency of thermal energy input from
supernova explosion to be $\epsilon < 0.01$.  This is significantly
smaller than both the thermalization efficiency estimated for M82
($0.3\le \epsilon\le 1$) by Strickland \& Heckman (2009) and those
found in numerical simulations (e.g.\ $\epsilon \approx 0.2$ from
Thornton \etal\ 1998).  We note, however, that the model is only
compatible with the observations for a special value of $\theta_0$ in
each case.  A significant fraction of the allowed $\theta_0$ would
yield a steeper velocity gradient, increasing velocity at larger
$z$-height, that is incompatible with the expectations of the model.
Such discrepancy suggests that if the gas is being accelerated, then
additional forces are necessary.

Finally, we consider the constraints for the power-law acceleration
field.  We have shown that the inferred $v(z)$ for different allowed
$\theta_0$ under an acceleration scenario requires that either the
superwinds are launched close to the star-forming disk with $z_{\rm
  min}=1$ kpc but a steeply declining acceleration $a(z)\propto
z^{-3}$ or they are launched at large distances with $z_{\rm min}\apg
4$ kpc and $a(z) \propto z^{-1.5}$.  The latter applies for most
allowed $\theta_0$.  Other mechanisms for lauching super-galactic
winds include cosmic ray pressure (e.g.\ Pfrommer \etal\ 2007; Everett
\etal\ 2008), but whether or not cosmic rays can explain a lauch
$z$-height beyond 4 kpc is unclear.

In summary, we have considered in detail the scenario of accelerated
outflows for explaining the observed gas kinematics of strong Mg\,II
absorbers.  We have derived the first empirical constraints for
super-galactic winds at $z_{\rm gal} \apg 0.5$ that can be compared
directly with model predictions.  Our analysis has uncovered an
interesting parameter space that is largely incompatible with current
models for driving accelerated outflows on galactic scales.
Decelerated outflows can explain the observed velocity field, but the
inferred velocity gradients (see Table 3) are significantly steeper
than expected from momentum-driven wind models (e.g.\ Dijkstra \&
Kramer 2012).  

A competing scenario is that some or all of the absorbing gas
originates in infaling halo gas or a larger co-rotating disk.
Unfortunately given the large degree of freedom in the spatial
distribution and trajectory of infalling clouds, we cannot place
meaningful constraints on the infall velocity field.  On the other
hand, Lanzetta \& Bowen (1992) have demonstrated that absorption-line
systems tracing co-rotating disks are expected to exhibit an
``edge-leading'' profile with the peak absorbing component occuring at
the largest velocity offset and declining absorbing strength toward
the systemic velocity of the absorbing galaxy.  Given that none of the
galaxy-absorber pairs in our sample shows such edge-leading profiles,
we can rule out the scenario of the absorbing gas following an
organized rotation motion around the central galaxy.


Lastly, given available observations of QSO absorption line systems at
low redshifts (the COS-Halos survey -- e.g.,
\citealt{tumlinson2012a}), we expect that a similar analysis like the
one presented in this paper can be conducted for a larger sample of
galaxies for which HST images are available.  In particular, the UV
spectra obtained by the Cosmic Origins Spectrograph cover both high-
and low-ionization transitions (e.g., Si\,II, C\,IV, and O\,VI) which
offer additional constraints for the kinematics of the hot
circumgalactic medium over a range of stellar masses and star
formation histories.

\section*{Acknowledgments}

We thank C.\ Steidel, R.\ Trainor, and Y.\ Matsuda for helpful
discussions during the early stages of this project.  We thank O.\
Agertz, N.\ Gnedin, A.\ Kravtsov, M.\ Rauch, G.\ Rudie, W.\ Sargent,
and A.\ Wolfe for helpful comments on an earlier version of the paper.
We also thank C.\ Steidel for providing the optical spectrum of galaxy
B in our study.  JRG gratefully acknowledges the financial support of
a Millikan Fellowship provided by Caltech and of a Grant-In-Aid of
Research from the National Academy of Sciences, administrated by Sigma
Xi, The Scientific Research Society.


\label{lastpage}

\end{document}